# Agent-based model of information spread in social networks

D.V. Lande [a,b], A.M. Hraivoronska [a], B.O. Berezin [a]

[a] *Institute for Information Recording NASU, Kiev, Ukraine*
[b] *NTUU "Kyiv Polytechnic Institute", Kiev, Ukraine*

*We propose evolution rules of the multiagent network and determine statistical patterns in life cycle of agents – information messages. The main discussed statistical pattern is connected with the number of likes and reposts for a message. This distribution corresponds to Weibull distribution according to modeling results. We examine proposed model using the data from Twitter, an online social networking service.*

## Introduction

The flows of information have a strong influence on opinion formation and other processes in the society. Today social networks play a fundamental role as a medium for the information spread. These facts motivate to explore mechanisms of creation of information flows and influence on them. Dealing with this requires focusing attention on modeling and finding laws or patterns in the spread of information [1].

In this article we present an agent-based model of information spread. The agent in this model is an information message [2]. A message published in social network may cause different types of public reaction. This model involves types of reaction such as positive or negative comments, respect or protest (we will call it *like/dislike*); message may be shared or copied (*repost*); also one message may have a link to another one (*link*). The evolution of the agent is controlled by mentioned above types of reaction. The main attribute of the agent is "energy" (**E**); that is representation of current relevance of the message or a degree of interest to the topic of the message by people. Naturally, a positive reaction or appearance of link to the message cause increase of energy. In opposite way, energy decreases when the message gets negative feedback. Anyway, energy tends to decrease because information eventually becomes outdated.

## The agent specification

More precisely the rules of agent evolution are as follows. Each agent appears with the initial energy (**E₀**) and dies when its energy becomes 0. The energy varies during the agent's life cycle depending on the types of reaction. Let us list them all and their impact on the energy:

- like: energy is incremented;
- dislike: energy is decremented;
- repost: energy is increased by 2;
- reference: energy is incremented;

In addition the energy is decremented at every time step (we consider the evolution in discrete time).

On the other hand, the more relevance of the message, the more likely people respond and express their opinion about information in this message. It is assumed the probability to get some response depends on current energy of agent. We introduce the probability of getting certain reaction for the agent with energy **E** as follows

$$p_{like}^{(E)} = p_{l_0}\varphi(E), \; p_{dislike}^{(E)} = p_{d_0}\varphi(E), \; p_{repost}^{(E)} = p_{r_0}\varphi(E).$$

We denote by $p_{l_0}, p_{d_0}, p_{d_0}$ initial parameters of the model, and by $\varphi$ some monotone nondecreasing function from $\mathbb{R}$ to [0, 1].

## The simulation of information spread

Earlier we introduced the evolution rules for the agent. The information flow consists of the set of such agents. We simulate the dynamics of the whole information flow as follows. At the initial time only one agent exists. New agents may appear in two ways. Firstly there is a probability of spontaneous generation ($p_s$). It means that new agent may appear with probability $p_s$ at every time step. Such appearance corresponds to the publishing new information by somebody. Secondly a copy of existing agent may be created (*repost*).

Here we describe the life cycle of one agent in terms of variation of its energy. Let $\varepsilon_t$ denote the value of energy at time $t$. Suppose $\delta_t$ is the random variable such that

$$P(\delta_t = 2|\varepsilon_t = E) = p_{like}^{(E)} \, p_{repost}^{(E)},$$

$$P(\delta_t = 1|\varepsilon_t = E) = \left(1 - p_{like}^{(E)}\right) p_{repost}^{(E)},$$

$$P(\delta_t = 0|\varepsilon_t = E) = p_{like}^{(E)} \left(1 - p_{repost}^{(E)}\right),$$

$$P(\delta_t = -1|\varepsilon_t = E) = \left(1 - p_{like}^{(E)}\right)\left(1 - p_{repost}^{(E)}\right).$$

Let us denote $P_\Delta^{(E)} = P(\delta = \Delta | \varepsilon = E)$. Then we have

$$\varepsilon_{t+1} = \varepsilon_t + \delta_t.$$

It follows that we can consider a change of energy as the random walk on $\{0, 1, 2, \dots, E_0, \dots\}$ with transition probabilities

$$p_{ij} = \begin{cases} P_{j-i}^{(i)}, (j-i) \in \{-1, 0, 1, 2\} \text{ и } i > 0, \\ 1, i = j = 0, \\ 0, otherwise. \end{cases}$$

In other words the stochastic sequence $(\varepsilon_0, \varepsilon_1, \dots, \varepsilon_t, \dots)$ is a Markov chain with transition probabilities $p_{ij}$. A state diagram for this Markov chain is shown on Figure 1, using a directed graph to picture the state transitions.

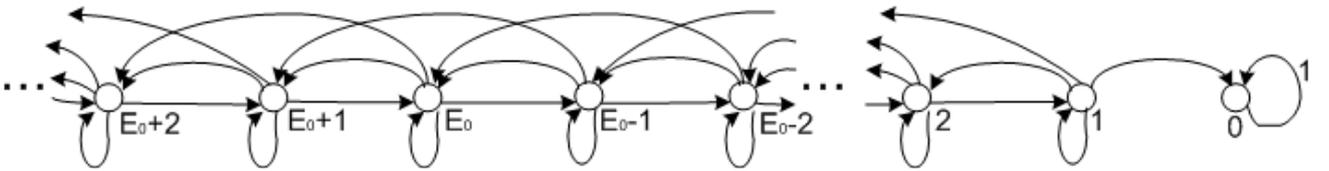

Figure 1. A state diagram for Markov chain. States represent energy of an agent

The random walk of energy is useful approach to analysis of properties of the model.

## Model results

Now let us consider the statistical distribution of likes and reposts for messages in the information flow. Note that we can find the probability to get n likes for one agent according to the above theoretical approach.

Suppose an agent gets like at time t; then $\delta_t \in \{0,2\}$, otherwise $\delta_t \in \{-1,1\}$. Denote by $(\Delta'_1, \ldots, \Delta'_{T_{max}})$ any vector such that $\Delta'_t \in \{0,2\}$, if $t = t_1, \ldots, t_n$ and $\Delta'_t \in \{-1,1\}$ otherwise for $0 < t_1 < \cdots < t_n < T_{max}$. It is easily proved that

$$P\{agent\ get\ n\ likes\} = \sum_{t_1 < \cdots < t_n} \sum_{(\Delta'_1, \ldots, \Delta'_{T_{max}})} \prod_{i=1}^{T_{max}} P_{\Delta'_i}^{(E_0 + \Sigma_{j=1}^{i-1} \Delta'_j)}.$$

Data generated by the model is illustrated in Figure 2.

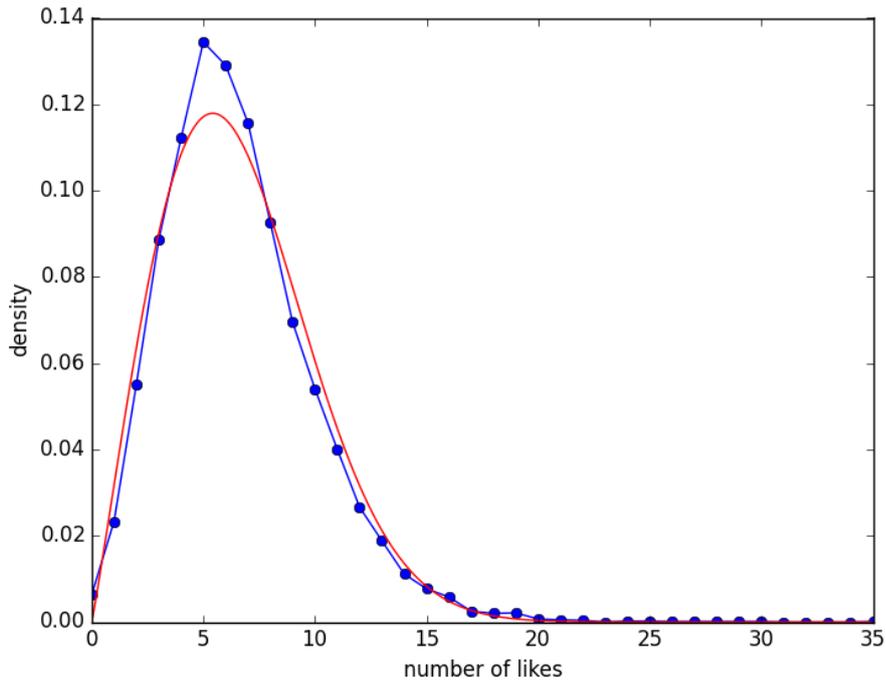

Figure 2. Distribution of likes generated by model

The frequency distribution of likes (blue line with dots) increases at first, and then decreases. It looks like a density function of the Weibull distribution [4]

$$f(x) = \begin{cases} \frac{k}{\lambda}\left(\frac{x}{\lambda}\right)^{k-1} e^{-\left(\frac{x}{\lambda}\right)^k}, & x \geq 0 \\ 0, & x < 0. \end{cases}$$

In Figure 2 a density function of the Weibull distribution with the shape parameter $k = 2.1$ and the scale parameter $\lambda = 7.4$ is shown (red line). We get this density function as an approximation for the frequency distribution of likes using the method of least squares.

The frequency distribution of reposts and its approximation are shown in Figure 3. Here a density function of the Weibull distribution has the shape parameter $k = 1.7$ and the scale parameter $\lambda = 4.6$.

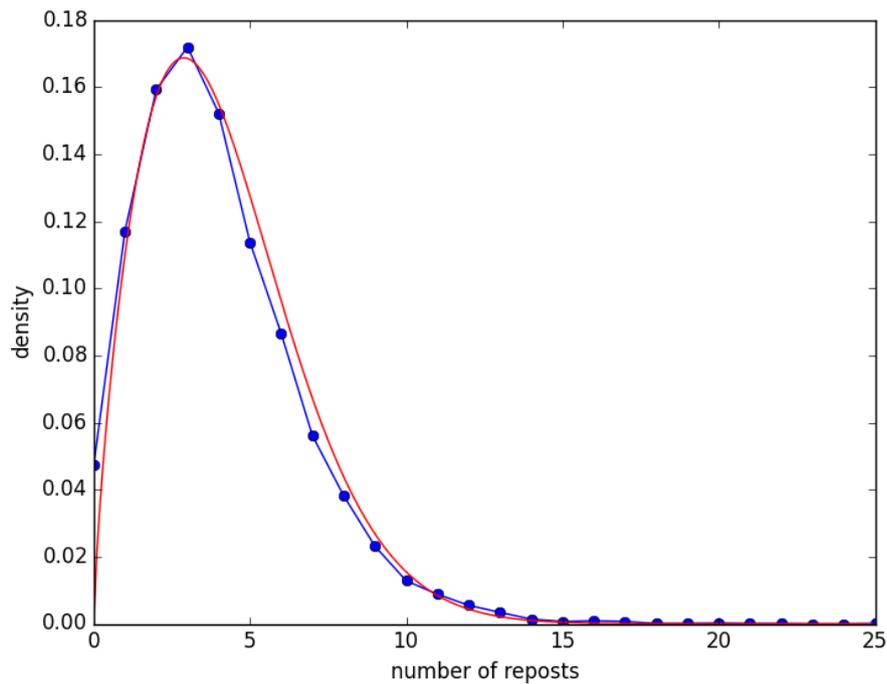

Figure 3. Distribution of reposts generated by model

**Information flows in social networks**

We study life cycles of news publications in Twitter and compare results with output produced by the model. Data about increase of likes and retweets for special information messages were collected [3]. We found that distributions of likes and retweets from a real social network fit Weibull distribution similarly to the model (Figure 5 and Figure 6). The shape parameter coincides with good accuracy in both situations.

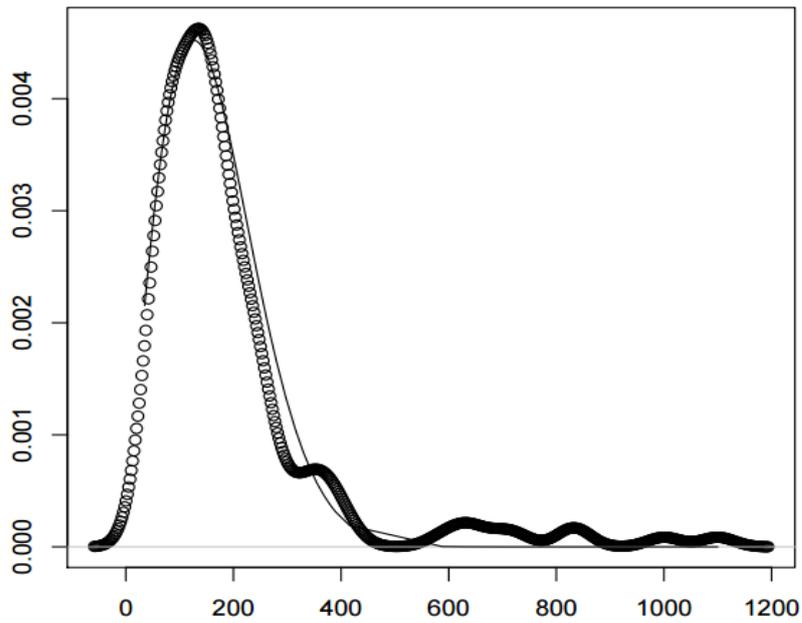

Figure 5. Distribution of likes from Twitter

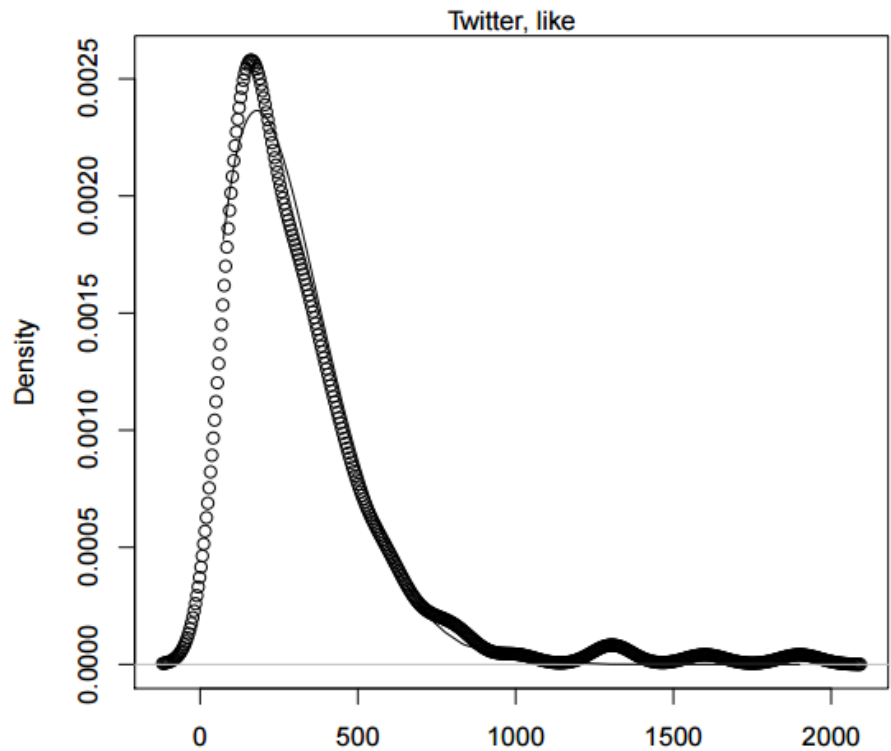

Figure 6. Distribution of retweets from Twitter

We developed a computer program using R programming language for analyzing statistical data. Processing was carried out in three following steps.

At first step the program detected increases in number of retweets for one user in online mode. For example, messages of New York Times newspaper were scanned every 15 minutes.

At second step the program treated data accumulated at the first step. We applied Weibull distribution for data approximation, so the scale parameter and the shape parameter were calculated. In addition, we estimated the rate of increase for number of retweets.

At third step all gathered data were stored in the external data base for future analysis.

To summarize, we collected texts of each message, timestamps for messages, scale and shape parameters, and numerous graphs. These graphs represent number of likes and retweets, the rate of growth for number of likes and retweets, and approximation for number of likes and retweets with Weibull distribution.

## Conclusion

We constructed agent-based model of message life cycle in social networks.

The statistical pattern for number of likes and reposts for information messages was found. Distribution of likes and reposts satisfy Weibull distribution according to modeling results. Model output is quite similar to the results from the real social network. It follows that the statistical pattern exists in real social networks and the model captures this pattern.

Findings described in this article can be useful for future studying of information spread in social networks. Also the presented results can be applied to detecting anomalies in a life cycle of information messages.

1. Dodonov A.G., Lande D.V., Prischepa V.V., Putiatin V.G. Competitive intelligence in computer networks . – K .: IPRI NAS of Ukraine, 2013. – 248 p.
2. Hraivoronska A.M., Lande D.V. Agent-based approach to investigating dynamics of the information flows // System analysis and information technologies: Materials of the 17[th] International scientific conference SAIT 2015, Kiev, 22-25 June 2015 / UNK IASA NTUU "KPI". – K .: ESC IASA NTUU "KPI", 2015. – C. 62-63.
3. Li R., Lei K.H., Khadiwala R., Chang K.C. TEDAS: A Twitter-based Event Detection and Analysis System  // Data Engineering (ICDE), 2012 IEEE 28th International Conference, 2012. — P. 1273-1276.
4. Neshitoy V.V. Mathematical and statistical analysis techniques in library and informationtional activities. – Minsk: Belarus. State. Univ. of Culture and Arts, 2009. – 203 p.